\documentclass[a4paper,11pt]{article}
\usepackage{atsirk}
\usepackage[cp1251]{inputenc}
\usepackage[russian]{babel}
\usepackage{graphicx}
\usepackage{amsmath}
\usepackage{amssymb}
\usepackage{amstext}
\begin{document}
\makeatletter
\renewcommand{\@oddhead}{}
\renewcommand{\@evenhead}{}
\renewcommand{\@evenfoot}{\hbox to \textwidth
  {Astron. Tsirkulyar No.~16\ldots\hfil\thepage\hfil May 2015}}
\renewcommand{\@oddfoot}{\hbox to \textwidth
  {Astron. Tsirkulyar No.~16\ldots\hfil\thepage\hfil May 2015}}
\renewcommand{\figurename}{Figure}
\renewcommand{\tablename}{Table}
\makeatother
{ISSN 0236-2457}

{\large\bf
\centerline{ASTRONOMICHESKII TSIRKULYAR}}
\medskip
\hrule
\medskip
\large
\centerline{Published by the Eurasian Astronomical Society}
\centerline{and Sternberg Astronomical Institute}
\medskip
\hrule
\medskip
\centerline{No.16\ldots, 2015 May \dots}
\medskip
\hrule
\bigskip
\centerline{\textbf{THE FAINT YOUNG SUN PARADOX IN THE CONTEXT}}
\centerline{\textbf{OF MODERN COSMOLOGY}\footnote{
Talk presented at the Interdisciplinary Colloquium
on Cosmic Factors of Evolution of the Biosphere and Geosphere,
Moscow, May 21--23, 2014.}}
\medskip
\centerline{\textbf{Yu.V. Dumin${}^{*,**}$}}
\smallskip
\centerline{\textit{${}^{*}$P.K.~Sternberg Astronomical Institute
            of M.V.~Lomonosov Moscow State University,}}
\centerline{\textit{Universitetskii prosp.\ 13, 119234, Moscow, Russia}}
\smallskip
\centerline{\textit{${}^{**}$Space Research Institute
            of Russian Academy of Sciences,}}
\centerline{\textit{Profsoyuznaya str.\ 84/32, 117997, Moscow, Russia}}
\smallskip
\centerline{\textit{E-mail: dumin@yahoo.com, dumin@sai.msu.ru}}
\smallskip
\centerline{\small Received May \dots , 2015}
\smallskip

\textbf{Abstract.}
The Faint Young Sun Paradox comes from the fact that solar luminosity
$ (2{\div}4){\cdot}10^9 $~years ago was insufficient to support the Earth's
temperature necessary for the efficient development of geological and
biological evolution (particularly, for the existence of considerable
volumes of liquid water).
It remains unclear by now if the so-called greenhouse effect on the Earth
can resolve this problem.
An interesting alternative explanation was put forward recently by
M.~K{\v{r}}{\'{\i}}{\v{z}}ek (New Ast. 2012, {\bf 17}, 1), who
suggested that planetary orbits expand with time due to the local Hubble
effect, caused by the uniformly-distributed Dark Energy.
Then, under a reasonable value of the local Hubble constant,
it is easy to explain why the Earth was receiving an approximately
constant amount of solar irradiation for a long period in the past
and will continue to do so for a quite long time in future.

\section*{Introduction}

The Faint Young Sun Paradox was recognized in the late 1950s, when
the sufficiently accurate models of stellar evolution were constructed.
As a result, it was found that luminosity of the Sun should change
considerably at the time scale of geological and biological evolution
of the Earth.
For example, the well-known monograph~[1] stated that the luminosity
increased by 1.6~times for the period of $ 5{\cdot}10^9 $~years.
Although a number of  subsequent models gave a somewhat less variation
in the luminosity ($ 30{\div}40\% $), the problem persists.
Namely, the average Earth's temperature $ 2{\cdot}10^9 $~years ago
is predicted to be so low that almost all water would be frozen.
On the other hand, a lot of geological and planetological evidences
suggest that there were extensive volumes of liquid water on the Earth
$ (3{\div}4){\cdot}10^9 $~years ago.
Besides, the liquid water would be necessary for the emergence of life
in the same period of time.

The commonly-used approach to resolve the Faint Young Sun Paradox,
widely exploited since 1970s~[2, 3], is the greenhouse effect,
\textit{i.e.}, keeping the infrared radiation emitted from the Earth surface
by the atmosphere.
The efficiency of such process strongly depends on the atmospheric
chemical composition (particularly, the minor constituents CO$_2$,
CH$_4$, NH$_3$, \textit{etc.}).
Since their concentration in the early atmosphere is unknown,
there is much uncertainty in the theoretical models.
In general, it remains unclear by now if the greenhouse effect can
be sufficiently efficient to compensate the reduced solar luminosity
in the past~[4].

\section*{Expansion of Planetary Orbits}

An interesting alternative idea for resolution of the Faint Young
Sun Paradox was suggested recently by M.~K{\v{r}}{\'{\i}}{\v{z}}ek
and L.~Somer~[5, 6].
They believe that the increasing solar luminosity could be compensated
by the expansion of the Earth's orbit, which is presumably caused by
the local Hubble effect associated with the uniformly-distributed
Dark Energy (or $ \Lambda $-term).

In fact, the question if planetary orbits experience the universal
cosmological expansion was posed by G.C.~McVittie in the early 1930's~[7],
and a quite large number of researchers worked on this problem in the
subsequent few decades (\textit{e.g.}, review~[8]).
Although the most of them concluded that Hubble expansion should be
strongly suppressed or absent at all at the sufficiently small scales,
there is no well-established criterion for such suppression.
Besides, some of the proposed criteria strongly contradict each other.
For example, observers usually assume that Hubble expansion should be
absent in the gravitationally-bound systems, \text{i.e.} in the regions
of \emph{enhanced density};
while the well-known Einstein--Straus model~[9] claims that Hubble
expansion is absent in the \emph{empty} local neighborhood of the central
body and is restored as the mass density increases up to the mean
cosmological value.
Moreover, the most of theoretical treatments are not applicable to
the case of cosmological models dominated by the Dark Energy,
which is assumed to be perfectly uniform and present everywhere~[10].

Therefore, the hypothesis by M.~K{\v{r}}{\'{\i}}{\v{z}}ek and L.~Somer
seems to be quite reasonable and, from our point of view, deserves
a more detailed consideration.
Of particular importance is the question if the local Hubble
parameter~$ H_0^{\rm (loc)} $ equals the global one and, if not, what is
its value?

To get some estimates, let us consider the solar luminosity increasing
linearly with time:
\begin{equation}
L(t) = L_0 + ( \Delta L / \Delta T ) \, t \, ,
\label{eq:Lumin}
\end{equation}
where
$ L_0 $~is the present-day luminosity,
$ \Delta T = 5{\cdot}10^9 $~yr, and
$ \Delta L / L_0 = 0.3, \, 0.4, \, 0.5, \, 0.6 $, which covers the entire
range of the solar models discussed in the literature.

Next, we assume that the mean radius of the Earth's orbit gradually increases
with time according to the Hubble law:
\begin{equation}
dR/dt = H_0^{\rm (loc)} R \, .
\label{eq:Hubble}
\end{equation}
Assuming for simplicity that $ H_0^{\rm (loc)} \! = {\rm const} $
(which should be a reasonable approximation when the Dark Energy dominates),
equation~(\ref{eq:Hubble}) can be easily integrated:
\begin{equation}
R(t) = R_0 \exp \! \big[ H_0^{\rm (loc)} t \big] \, .
\label{eq:Rad_t}
\end{equation}

Then, solar irradiation (the energy flux density) at the Earth,
$ F \! = L / ( 4 \pi R^2 ) $,
will change with time as
\begin{equation}
F(t) = F_0 \Big[ 1 + ( \Delta L / L_0 ) ( t / \Delta T ) \Big]
  \exp \!\! \big[ \! - \!\! 2 ( H_0^{\rm (loc)} \! / H_0 ) ( H_0 t) \big] \, ,
\label{eq:Flux_t}
\end{equation}
where
$ H_0 \! \approx $ 70\,(km/s)/Mpc~$ \approx 0.071{\cdot}10^{-9} $\,yr$^{-1}$,
and time~$ t $ can be conveniently expressed in the units of $ 10^9 $\,yr.
A number of corresponding temporal dependences for different models of solar
evolution, characterized by $ \Delta L / L_0 $, and various values of
the local Hubble para\-meter~$ H_0^{\rm (loc)} $ are presented in Figure~1.

\begin{figure*}[t]
\begin{center}
\includegraphics[height=14.4cm]{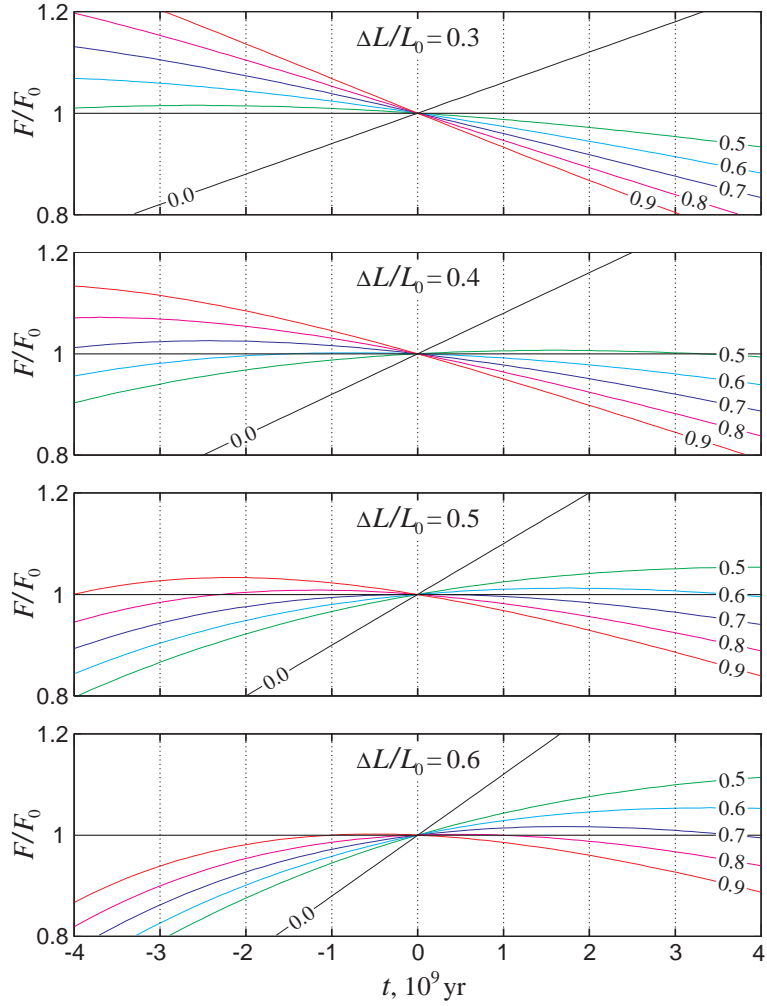}
\end{center}
\vspace*{-8ex}
\caption{Relative variation of the solar irradiance at the Earth
as function of time for a few solar models (four panels, from top to bottom)
at various values of the local Hubble constant:
$ H_0^{\rm (loc)} \! / H_0 = $ 0.5~(green curves), 0.6~(cyan), 0.7~(blue),
0.8~(magenta), and 0.9~(red).
Black inclined lines correspond to the case without cosmological correction,
$ H_0^{\rm (loc)} \! / H_0 = 0.0 \, $.}
\end{figure*}

\section*{Discussion}

To resolve the Faint Young Sun Paradox, we should seek for such
regimes of evolution when the relative flux density $ F(t)/F_0 $
deviates as small as possible (\textit{e.g.}, within a few percent)
from unity in the entire time interval from $ -(3{\div}4){\cdot}10^9 $\,yr
up to the present time, $ t=0 $.
So, the following principal conclusions can be derived from the analysis
of Figure~1:

{\bf 1.}~For the solar model with a relatively small variation in
the luminosity, $ \Delta L / L_0 = 0.3 $ (top panel), the most favorable
regime is achieved at $ H_0^{\rm (loc)} \! = 0.5 \, H_0 $.
This is approximately the case considered by M.~K{\v{r}}{\'{\i}}{\v{z}}ek
and L.~Somer~[5, 6], which provides a very stable solar irradiation in
the past and a quite gradual deviation in the future.

Unfortunately, this value of the local Hubble constant is poorly consistent
with the conjecture that just the Dark Energy is responsible for
the expansion of planetary orbits.
Really, let us assume that
(i)~the local Hubble expansion is produced only by the perfectly-uniform
Dark Energy, while the clumped substance manifests itself as Newtonian
forces; and
(ii)~the global Hubble expansion is formed both by the Dark Energy and
average contribution from the ordinary matter.
Then, it can be shown~[11] that the local Hubble parameter is related to
the global one as
\begin{equation}
\frac{H_0^{\rm (loc)}}{H_0} =
  {\left[ 1 + \frac{\Omega_{{\rm D}0}}{\Omega_{{\Lambda}0}} \, \right]}^{-1/2}
\!\! \approx \:
  1 - \frac{1}{2} \, \frac{\Omega_{{\rm D}0}}{\Omega_{{\Lambda}0}} \: ,
\label{eq:H_loc_glob}
\end{equation}
where $ \Omega_{{\Lambda}0} $ and $ \Omega_{{\rm D}0} $~are densities
of the Dark Energy ($\Lambda$-term) and ordinary substance
(for the most part, Dark Matter).
So, taking
$ \Omega_{{\Lambda}0} \! \approx 0.7 $ and
$ \Omega_{{\rm D}0} \! \approx 0.3 $, we get
$ H_0^{\rm (loc)} \! / H_0 \! \approx 0.8 $,
which is substantially different from the above-mentioned value~0.5\,.

{\bf 2.}~On the other hand, as is seen in the third panel of Figure~1,
the rate of local expansion $ H_0^{\rm (loc)} \!\! = 0.8 H_0 $ enables us
to get a sufficiently stable irradiation for the solar model with
an increased variability, $ \Delta L / L_0 \! = 0.5 $
(but the overall stability is not so good as in the first case,
especially, in the future period).
However, just the second case seems to be more consistent with
the cosmological argumentation.

{\bf 3.}~Anyway, it should be kept in mind that the greenhouse effect is
of considerable importance in the thermal history of the Earth's atmosphere.
So, the temporal variation of the solar irradiation cannot be immediately
confronted with geological and biological evidences.

\section*{References}

\hspace{6.5mm}

1. Schwarzschild~M.
\emph{Structure and Evolution of the Stars}
(Princeton, N.J.: Princeton Univ.\ Press, 1958).

2. Sagan~C. and Mullen~G.,
% Earth and Mars: Evolution of Atmospheres and Surface Temperatures
Science, \textbf{177}, 52 (1972).

3. Sagan~C.,
% Reducing Greenhouses and the Temperature History of Earth and Mars
Nature, \textbf{269}, 224 (1977).

4. Martens~P.C., Nandi~D., and Obridko~V.N.,
% Project SEE
VarSITI Newsletter, \textbf{1}, 2 (2014) \\
(http://www.varsiti.org).

5. K{\v{r}}{\'{\i}}{\v{z}}ek~M.,
% Dark Energy and the Anthropic Principle
New Ast., \textbf{17}, 1 (2012).

6. K{\v{r}}{\'{\i}}{\v{z}}ek~M. and Somer~L.,
% Manifestations of Dark Energy in the Solar System
Grav.\ \& Cosmol., \textbf{21}, 59 (2015).

7. McVittie~G.C.,
% The Mass-Particle in an Expanding Universe
MNRAS, \textbf{93}, 325 (1933).

8. Bonnor~W.B.,
% Local Dynamics and the Expansion of the Universe
Gen.\ Rel.\ \& Grav., \textbf{32}, 1005 (2000).

9. Einstein~A. and Straus~E.G.,
% The Influence of the Expansion of Space on the Gravitation
% Fields Surrounding the Individual Stars
Rev.\ Mod.\ Phys., \textbf{17}, 120 (1945).

10. Dumin~Yu.V.,
% A New Application of the Lunar Laser Retroreflectors:
% Searching for the Local Hubble Expansion
Adv.\ Space Res., \textbf{31}, 2461 (2003).

11. Dumin~Yu.V.,
% Testing the Dark-Energy-dominated Cosmology by
% the Solar-System Experiments.
in: \emph{Proc. 11th Marcel Grossmann Meeting on General Relativity},
p.1752 (Singapore: World Scientific, 2008).

\newpage

\centerline{\textbf{ПАРАДОКС ТУСКЛОГО МОЛОДОГО СОЛНЦА В КОНТЕКСТЕ}}
\centerline{\textbf{СОВРЕМЕННОЙ КОСМОЛОГИИ}}
\medskip
\centerline{\textbf{Ю.В. Думин${}^{*,**}$}}
\centerline{\textit{${}^{*}$Государственный астрономический институт
            им.~П.К.~Штернберга}}
\centerline{\textit{Московского Государственного Университета
            им.~М.В.~Ломоносова,}}
\centerline{\textit{Университетский просп., 13, 119234, Москва, Россия}}
\smallskip
\centerline{\textit{${}^{**}$Институт космических исследований
            Российской Академии Наук}}
\centerline{\textit{Профсоюзная ул., 84/32, 117997, Москва, Россия}}
\smallskip
\centerline{\textit{E-mail: dumin@yahoo.com, dumin@sai.msu.ru}}
\smallskip

\textbf{Резюме.}
Парадокс ``тусклого молодого Солнца'' связан с тем, что его светимость
$ (2{\div}4){\cdot}10^9 $~лет назад была недостаточной для поддержания
на Земле температуры, необходимой для эффективного протекания
геологической и биологической эволюции (в частности, для существования
значительного количества воды в жидком состоянии).
До сих пор остается неясным, может ли ``парниковый'' эффект в
достаточной степени решить эту проблему.
Интересное альтернативное объяснение было недавно предложено
М.~Кризеком (M.~K{\v{r}}{\'{\i}}{\v{z}}ek, New Ast. 2012, {\bf 17}, 1),
который предположил, что орбиты планет увеличиваются с течением
времени за счет локального эффекта Хаббла, связанного с
однородно распределенной ``темной энергией''.
Тогда, при разумном значении локальной постоянной Хаббла легко объяснить,
почему Земля получала приблизительно постоянную плотность потока
солнечного излучения на протяжении длительного периода в прошлом
и будет получать ее еще достаточно долгое время в будущем.
\end{document}